# The New Horizons Pluto Kuiper belt Mission: An Overview with Historical Context

S. Alan Stern[a]

[a] Southwest Research Institute, 1050 Walnut St., Suite 400, Boulder, CO 80302

**Abstract**

NASA's New Horizons (NH) Pluto-Kuiper belt (PKB) mission was launched on 19 January 2006 on a Jupiter Gravity Assist (JGA) trajectory toward the Pluto system for a 14 July 2015 closest approach; Jupiter closest approach occurred on 28 February 2007. It was selected for development on 29 November 2001 following a competitive selection resulting from a NASA mission Announcement of Opportunity. New Horizons is the first mission to the Pluto system and the Kuiper belt; and will complete the reconnaissance of the classical planets. The ~400 kg spacecraft carries seven scientific instruments, including imagers, spectrometers, radio science, a plasma and particles suite, and a dust counter built by university students. NH will study the Pluto system over a 5-month period beginning in early 2015. Following Pluto, NH will go on to reconnoiter one or two 30-50 kilometer diameter Kuiper belt Objects (KBOs) if NASA approves an extended mission. New Horizons has already demonstrated the ability of PI-led missions to be launched to the outer solar system. As well, the mission has demonstrated the ability of non-traditional entities, like APL and SwRI to explore the outer solar system, giving NASA new programmatic flexibility and enhancing the competitive options when selecting outer planet missions. If successful, NH will represent a watershed development in the scientific exploration of a new class of bodies in the solar system—dwarf planets, of worlds with exotic volatiles on their surfaces, of rapidly (possibly hydrodynamically) escaping atmospheres, and of giant impact derived satellite systems. It will also provide the first dust density measurements beyond 18 AU, cratering records that shed light on both the ancient and present-day KB impactor population down to tens of meters, and a key comparator to the puzzlingly active, former dwarf planet (now satellite of Neptune) called Triton which is as large as Eris and Pluto.

# 1. Mission Overview

**Background.** New Horizons is a flyby reconnaissance mission to provide the first in situ exploration of the Pluto system and Kuiper Belt Objects (KBOs). It is the first mission in NASA's New Frontiers series of medium-class, robotic planetary exploration missions. The New Horizons flight system consists of a single, ~400 kg spacecraft featuring highly redundant subsystems and 7 scientific instruments. The spacecraft is powered by a Radioisotope Thermoelectric Generator (RTG).

The top level goals of the New Horizons science mission are, in priority order:

- ➢ Reconnoiter the Pluto system for the first time.
- ➢ Sample the diversity of KBOs by making one or more KBO flybys after the Pluto system flyby.
- ➢ Obtain Jupiter system science during the Jupiter Gravity Assist.

The specific scientific measurement objectives of the mission were developed by NASA's Outer Planets Science Working Group (OPSWG, S.A. Stern chair) in 1992 and slightly refined and then re-ratified by the Pluto Kuiper Express (PKE, J.I. Lunine, chair) mission Science Definition Team (SDT) in 1996 (Lunine et al. 1996). These objectives were adopted by NASA for the mission AO that led to the selection of New Horizons.

The suite of New Horizons mission science objectives are ranked in three categories, called Group 1, Group 2, and Group 3. This categorization was first developed by OPSWG (then denoted Group IA, Group IB, and Group IC) and reaffirmed by the PKE SDT. Group 1 objectives represent an irreducible floor for the mission science goals at the Pluto system. Group 2 goals add depth and breadth to the Group 1 objectives and are termed highly desirable. The Group 3 objectives add further depth and are termed desirable, but of distinctly lower priority than the Group 2 objectives. The various objectives can be summarized as follows:

**Table 1. New Horizons Pluto-Charon and KBO Measurement Objectives**

| |
|---|
| **Group 1: Required** |
|   Characterize the global geology and morphology of Pluto and Charon |
|   Map surface composition of Pluto and Charon |
|   Characterize the neutral atmosphere of Pluto and its escape rate |
| **Group 2: Highly Desired** |
|   Characterize the time variability of Pluto's surface and atmosphere |
|   Image Pluto and Charon in stereo |
|   Map the terminators of Pluto and Charon with high resolution |
|   Map the surface composition of selected areas of Pluto and Charon at high resolution |
|   Characterize Pluto's ionosphere and solar wind interaction |
|   Search for neutral atmospheric species including H, $H_2$, HCN, & $C_xH_y$, and other hydrocarbons and |
|   Search for an atmosphere around Charon |
|   Determine bolometric Bond albedos for Pluto and Charon |
|   Map the surface temperatures of Pluto and Charon |
| **Group 3: Desirable** |
|   Characterize the energetic particle environment of Pluto and Charon |
|   Refine bulk parameters (radii, masses, densities) and orbits of Pluto and Charon |
|   Search for additional satellites and rings |

Each of the objectives named above was defined by the SDT in significantly more detail, giving measurement requirements that included resolutions, SNRs, dynamic ranges, etc., as appropriate. These will be described and discussed in the accompanying article by Young et al.

Since Pluto's small moons Nix and Hydra were not known at the time these objectives were constructed, the objectives do not rank the desirability of studying small satellites prior to any encounter. New Horizons, however, is in the enviable position of knowing about and being able to plan science observations for Nix and Hydra with over 9 years advance notice. We as a mission are treating the compositional and geologic mapping of these bodies as an additional objective.

This article provides an overview of the mission, describing its history, its goals, its architecture and development, and its flight in summary form. Other articles in this volume provide more in-depth background regarding the mission, the mission science, and the spacecraft and its instrument payload.

## 2. Pluto Mission Background Studies and Selection

In this section I will briefly recapitulate the relevant history of Pluto mission studies. I begin with the Voyager mission and work forward monotonically in time through the many studies of the 1990s, to the ultimate call for mission proposals in early 2001 and the selection of New Horizons at the end of that year.

**Voyager Pluto**. As did the never-built Grand Tour mission of the 1970s, NASA's Voyager 1 and 2 outer planets reconnaissance flyby missions included an option for Voyager 1 to fly from Saturn in 1980 to a late 1980s Pluto flyby. This option, however, was mutually exclusive with Voyager 1 making a close flyby of Saturn's large and complex, atmosphere-laden moon Titan during its late-1980 exploration of the Saturn system. Owing in part to the lower risk of the Titan flyby than a long cruise to Pluto, and also the higher scientific priority at the time of Titan, the Pluto option was not exercised. Of course, at the time this decision was made, Pluto's atmosphere, small satellites, complex surface composition, and the entire Kuiper belt all remained undiscovered. By the time of the 1989 Voyager 2 flyby of Pluto-analog Triton, Pluto's richness and context was beginning to be understood. This, combined with the fascinating results of Voyager 2's Triton flyby that included a pathologically young surface, active geysers, and an atmosphere, motivated interest, particularly in a handful of young planetary scientists, to successfully appeal to NASA in 1989 to begin Pluto mission studies.

**Pluto Mission Studies.** NASA began studying dedicated Pluto flyby reconnaissance missions at the behest of the scientific community in 1989. The first such study (eventually dubbed "Pluto-350") was undertaken as a part of the Discovery Program Science Working Group (DPSWG) in 1989-1990. The study scientists for this effort were S.A. Stern and F. Bagenal; the study manager was R. Farquhar. The concept for this study was to send a "minimalist" scientific payload to Pluto-Charon (Pluto's smaller moons Nix and Hydra were only discovered in 2005); the Kuiper belt was then undiscovered and not a part of the mission study. The resulting spacecraft (Farquhar & Stern 1990) was a 350-kg vehicle with four instruments (imager, UV spectrometer, radio science, and a plasma package). At the time of this study, a four-instrument spacecraft weighing half what Voyager did was considered controversial, both in terms of its small scope and its perceived high risk.

Shortly after the Pluto-350 study, NASA next studied flying a much larger, Cassini-class Mariner Mark II mission to Pluto. This mission, though much more costly, was perceived to have lower risk and a broader scientific potential. Notably, it would have replaced the Cassini Huygens Titan entry probe with a short-lived, deployable second flyby spacecraft designed to fly over Pluto's far hemisphere some 3.2 days (one Pluto half-rotation) before or after the mother ship. This mission was adopted as a high priority in the Solar System Exploration Subcommittee (SSES) 1990s planetary exploration plan derived in a "community shoot out" meeting in February 1991. Following this, NASA's Solar System Exploration Division (under the direction of W. Huntress) formed the Outer Planets Science Working Group (OPSWG; S.A. Stern, chair) to shape the mission's scientific content, document its rationale, and prepare for an instrument selection process by the mid-1990s. By 1992, OPSWG had completed most of its assigned mission study support tasks. Owing to tightening budgets at NASA, OPSWG also was asked to debate the large Mariner Mark II versus the much smaller Pluto-350 mission concepts. It did. In early 1992, OPSWG selected Pluto-350 as the more pragmatic choice.

However, in the late spring of 1992, a new, more radical mission concept called Pluto Fast Flyby (PFF) was introduced by JPL's R. Staehle as a "faster, better, cheaper" alternative to the Mariner Mark II and Pluto 350 Pluto mission concepts. As initially conceived, PFF was to weigh just 35-50 kg and carry only 7 kg of highly miniaturized (then non-existent) instruments, and fly two spacecraft to Pluto for <$500M. PFF caught the attention of then NASA Administrator D. Goldin, who directed all Pluto-350 and Mariner Mark II work to cease in favor of PFF. PFF would have launched its two flyby spacecraft on Titan IV-Centaur launchers; these low-mass spacecraft would have shaved the Pluto-350 and Mariner Mark II flight times from 12-15 years down to 7 or 8 years. Like Mariner Mark II and Pluto-350, PFF involved RTG power and Jupiter Gravity Assists (JGAs). The heavier missions also involved Earth and Venus gravity assists on the way to Jupiter. All of these mission concepts were developed by JPL mission study teams.

Shortly after PFF was introduced, however, it ran into problems. One was mass growth, which quickly escalated the flight system to the 140 kg class with no increase in mission payload mass. A second issue involved cost increases, largely due to a broad move within NASA to include launch vehicle costs in mission cost estimates; since two Titan IV launchers alone cost over $800M, this pushed PFF to well over $1B. A third issue was the turmoil introduced into NASA's planetary program by the loss of the Mars Observer in 1993. These events began to sour then NASA Administrator Goldin on PFF. Cost concerns subsequently caused the PFF concept to be cut back to one spacecraft.

Following this, OPSWG chair Stern attempted to gain European and Russian collaboration in the mission to reduce cost so that a new start could be afforded. European interest was generally lukewarm. However, Russian interest was stronger. A concept emerged between Stern and Russia's director A. Galeev of the IKI space research center in Moscow that a Russian Proton launch vehicle would loft PFF, saving NASA the ~$400M cost of the Titan IV launch. In it for Russia would be a probe, called a Drop Zond, which would enter Pluto's atmosphere to make measurements of it and imagery before an impact on Pluto. When Russia later (1995) demanded to be paid for this launch, Germany's Max Plank Institute for Planetary Physics offered to pursue German national funding for the Russian launch; the plan of the German scientists was to pay Russia for the Proton launch (~$30M at that time) in exchange for NASA accommodating a second probe on PFF which would impact Jupiter's moon Io during the JGA encounter.

PFF was never started into development owing to higher priorities within NASA for then Administrator Goldin. During 1994-1995, Mr. Goldin directed a series of studies to determine if PFF could fly without any nuclear power to Pluto, and also whether it could be launched on a small launcher (i.e., a Delta II) to reduce cost. These studies were widely considered in the outer planets community to be diversionary tactics by Mr. Goldin, who was perceived as not being able to cancel the Pluto effort but was unwilling to start it. Nonetheless, JPL carried the requested studies over a period of about a year; they concluded that although a slow (12-15 year) Delta II launched mission was feasible, non-nuclear missions were either too risky (e.g., using battery power alone) or beyond the cost or technological capability of the era. During this same period, PFF did solicit, select, and fund the breadboard/brassboard development of a suite of competitively miniaturized imagers, spectrometers, and radio science, and plasma instruments.

Following on the rapidly expanding interest in the Kuiper belt by the mid-1990s, NASA directed JPL to reinvent PFF as Pluto Express (later named and more commonly known as Pluto-Kuiper Express or PKE). PKE was a single spacecraft PFF mission with a 175 kg spacecraft and a 9-kg science payload. It would have launched in the 2001-2006 JGA launch window. A Science Definition Team (SDT) chaired by J.I. Lunine was constituted in 1995 and delivered its report in 1996 for an anticipated instrument selection in 1996-1997. However, in late 1996 PKE mission studies were drastically cut back by Administrator Goldin and no instrument selection was initiated.

By 1999, however, NASA did release a solicitation for PKE instruments; proposals were due in March 2000. Many of the proposals received resulted from the PFF miniaturized instrument development program. These proposals were evaluated and ranked, but never selected. By September 2000, NASA cancelled PKE, still in Phase A, owing to mission cost increases which pushed the projected mission cost over the $1B mark.

Following this cancellation, intense scientific and public pressure caused then NASA Associate Administrator for Space Science E. Weiler to solicit mission proposals in 2001 for a Pluto Kuiper belt flyby reconnaissance mission. That solicitation and the resulting selection of New Horizons are discussed just below. For additional details about early Pluto mission studies the following references are recommended: Stern (1993), Terrile et al. (1997), and Stern & Mitton (2005).

**PKB Mission AO and Selection Process.** The PKB AO was announced in a NASA Press Conference on 20 December 2000 and released on 19 January 2001. The AO (NASA 2001) mandated a two-step selection process with initial proposals due 20 March 2001 (later extended to 6 April 2001). Following a down-select to two teams, Phase A studies would be performed with due dates in the August-September timeframe. Because no PI-led mission to the outer planets, nor any PI-led mission involving RTGs, had ever been selected, the AO was termed experimental by NASA, who made clear they were not obligating themselves to select any proposals at all.

The PKB AO (01-OSS-10) required responders to propose an entire PKB mission, to meet at least the detailed specifications of the Group 1 measurement objectives, to complete a Pluto flyby by 2020, to launch aboard a U.S. Atlas V or Delta IV launch vehicle, and to do so within a complete mission cost cap of $506M FY2001 dollars. Launch vehicle selection between the Atlas V and Delta IV was planned for 2002. Two spare Cassini-Galileo RTGs were made available for use to proposal teams, with associated costs of $50M and $90M (the latter with higher power).

Shortly after the 19 January 2001 AO release, on 6 February 2001, the then-new Bush Administration released its first budget, which cancelled PKB by not funding it in FY02 and future years. Within days, NASA announced the suspension of the PKB AO as well. However, intensive work on Capital Hill within the science community resulted in less than a week in a directive from the U.S. Senate to NASA to proceed with the AO so as not to limit Congressional authority to override the PKB cancellation decision.

Five proposals were turned in to NASA. I understand that three were apparently strong contenders and two were not. The contenders included two proposals from JPL (L. Soderblom and L. Esposito, PIs) and one from APL (S.A. Stern, PI). The Soderblom et al. proposal cleverly involved ion propulsion in order to remove the 2004-2006 JGA launch window constraint. The Esposito et al. and Stern et al. proposals both involved conventional propulsion JGA trajectories. Since my only first-hand knowledge regarding these proposals is with New Horizons, and since NASA does not release details on the proposals not selected, I will not describe these or the other submitted proposals further. I will however summarize the New Horizons mission as proposed.

The New Horizons team was formed by an agreement between PI Stern and APL Space Department Head Dr. Tom Krimigis that was made on 22 December 2000. The science team was formed from Stern's PKE PERSI instrument proposal team and Dr. Len Tyler's PKE radio science proposal team, plus about 5 additional scientists added from APL and other institutions to add scientific breadth for a full mission proposal. Dr. Andy Cheng was named the New Horizons project scientist. The Tyler et al. radio science team had been the only radio science proposal for the PKE AO and their participation was considered by PI Stern to be a key strategic element of a winning PKB proposal.

The first face-to-face meeting of the New Horizons science and spacecraft teams took place at APL on 8 January 2001. Payload selection was largely complete by 22 January, just after the PKB AO was released. The mission concept was to launch a small (400 kg class) flyby spacecraft based on heritage from APL's CONTOUR multi-comet flyby mission then in development for launch in 2002. The PKB spacecraft would be able to fly about 30 kg of instruments—far more than the 7 to 9 kg of the JPL PFF-PKE era. It also would include substantial avionics and propulsion system redundancy for the long voyage, and it would use the lower-power (and lower cost) of the two RTGs that NASA offered in the AO.

In the proposal, strong proposal emphasis was placed on reducing programmatic (i.e., cost and schedule) risk because (i) APL was viewed as a new entrant to outer planet missions and (ii) it was important to convincingly avoid the repeated cost escalations of 1990s Pluto study and mission development attempts at JPL. A very large, 48 Gbit solid state memory was proposed for the mission in order to allow the spacecraft to take maximum advantage of its time in the Pluto system (by contrast, the PKE mission planned an 8 Gbit memory). Finally, every effort was made to propose the earliest feasible launch and arrival; so we proposed that launch would be in December 2004, with a January 2006 backup JGA. The December 2004 launch would target a July 2012 arrival. Our PKB proposal was named New Horizons, after a long process of winnowing, on 5 February 2001. The name was meant to symbolize both the new scientific horizons inherent in the exploration of the Pluto system and the Kuiper belt, as well as the programmatic new horizons of PI-led outer planet missions. PI Stern commissioned planetary scientist and artist Dan Durda to provide a "2001-esque" Pluto flyby graphic that evoked a sense of new horizons. That image, with an as-launched New Horizons substituted for the 2001-era concept, is shown in Figure 1 below.

The New Horizons payload as proposed consisted of the following four instrument packages:

**PERSI**, a PKE-proposed instrument package consisting of the ALICE UV spectrometer and the RALPH multi-color imager/IR imaging spectrometer.
**REX**, an uplink radio science instrument with radiometer capabilities.
**LORRI**, a long focal length panchromatic CCD camera to provide 5x higher resolution imaging than RALPH could accomplish.
**PAM**, a plasma package consisting of both the SWAP high-sensitivity solar wind monitor to address Pluto atmospheric escape objectives and the PEPSSI energetic particle spectrometer adapted from the EPS instrument then in development for MESSENGER Mercury orbiter.

Table 2 provides some additional details on the payload as proposed. The article in this volume by Weaver et al. provides a more detailed overview of the as-launched scientific payload, which differs only in terminology from that described here.

Table 2. Proposed New Horizons Payload

| Instrument | Type | Sensor Characteristics | Builders |
|---|---|---|---|
| **PERSI** | Remote sensing suite | MVIC (panchromatic and four-color CCD imager, 0.4-1.0 microns, 20 microradians/pixel), LEISA (near infrared imaging spectrometer, wedged filter, 1.25-2.5 µm, R=600 for 2.1-2.25 microns and R=300 otherwise, 62 microradians/pixel), and ALICE (UV imaging spectrometer, 500-1850 Å, spectral resolution 3 Å, 5 milliradians/pixel) | Ball, SwRI, NASA/GSFC |
| **REX** | Uplink radio science, passive radiometry | Signal/noise power spectral density 55 db-Hz; ultrastable oscillator stability $1 \times 10^{-13}$ in 1 second samples. Disk-averaged radiometry to ±0.1 K. | Stanford, JHU/APL |
| **PAM** | Plasma & high energy particle spectrometers | SWAP (solar wind plasmas up to 6.5 keV, toroidal electrostatic analyzer and retarding potential analyzer), and PEPSSI (ions 1-5000 keV and electrons 20-700 keV, time-of-flight by energy to separate pickup ions) | SwRI, JHU/APL |
| **LORRI** | High resolution imager | Panchromatic, narrow angle CCD imager, 0.30-0.95 microns, 5 microradians/pixel | JHU/APL |

The objectives of this payload were to significantly exceed the minimum mission requirements laid out by the AO, to significantly exceed what PKE would have accomplished, but not to overburden the mission with a costly "Christmas tree" array of instruments incompatible with a highly cost constrained outer planets mission. Other instruments considered but not included in this payload for various reasons were a magnetometer, a plasma wave senor, a dust instrument for cruise science in the deep outer solar system and the Kuiper belt, bolometers, and a mass spectrometer.

PERSI and REX were termed the New Horizons "core payload" because they were sufficient to accomplish all of the Group 1 science which the PKB AO required proposers to meet. LORRI and PAM were termed the "supplementary payload" designed to add depth and breadth to what the core instruments could do; however, the supplementary payload was clearly stated to be descopable if technical or programmatic considerations forced cutbacks.

I now return to the proposal process. Proposals were turned in on 6 April 2001. After a two month technical and programmatic review process, NASA announced the selection of JPL's POSSE (Pluto Outer Solar System Explorer; L. Esposito, PI) and APL's New Horizons for Phase A studies and further competition on 6 June 2001. PI Stern was at a Kuiper Belt meeting in Paris and was informed by a phone message to call home to "Dr. Yung" (meaning CoI Leslie Young), who relayed to him that NASA had called with the selection news somewhat earlier in the day. A kickoff meeting for Phase A development was sponsored by NASA headquarters on 18 June 2001.

Both POSSE and New Horizons were funded by NASA at the $500,000 level for phase A studies that were to be due on 18 September 2001. Both teams contributed substantial internal resources to supplement the

NASA funding they received. The ground rules of the Phase A study were that the proposal teams could not augment their science or science teams, but were instead to provide additional engineering, cost, and schedule study to further flesh out their mission concepts. The 11 September 2001 terrorist attacks on New York and Washington, D.C. interceded in the final days of proposal preparation. Owing to the nationwide stoppage of air transport (including overnight mail), a shut down of government activities in central Washington, D.C. for several days, and the general national paralysis that ensued, NASA extended the proposal deadline to 25 September.

Formal oral briefings on the proposals to a NASA Concept Study Evaluation Review Board were held for New Horizons and POSSE on 17 and 19 October, 2001, respectively. Stern & and Cheng (2002) and Stern (2002) summarized the mission as proposed.

In parallel with the Phase A and proposal activities described above, the scientific community and the New Horizons team also undertook a difficult effort to put funding in place in the NASA budget for FY02's needed Phase B development. Had this not been done, any selection of a mission would have been moot, since no contract could be let to begin work, thereby ensuring that the 2004-2006 JGA launch window would not be met and no mission would be built (the next JGA window would not open until 2015). Ultimately, after much work and much intrigue, this effort succeeded with the Senate passage and House-Senate conference agreement of a NASA FY02 budget in early 2002 that included $30M in supplementary funding for PKB mission to initiate spacecraft and science instrument development as well as work toward launch vehicle procurement.

NASA selected New Horizons in mid-November. However, the formal announcement of this award was held up until 29 November. PI Stern was informed of the selection of New Horizons for Phase B development by a phone call from NASA's PKB Program Scientist, Dr. Denis Bogan, while he was at the annual AAS Division of Planetary Sciences meeting, which was held in New Orleans that year. A win party was held on Bourbon Street that night in the New Orleans French Quarter, but the details remain understandingly fuzzy.

## 3. Mission Development Overview

Initiating the development of New Horizons was difficult for a variety of reasons. To begin, NASA Associate Administrator for Space Science Weiler's selection letter to PI Stern pointed out that many obstacles the mission faced before it could be confirmed. Among these were a lack of funding or plan to fund after Phase B; the lack of a nuclear qualified launch vehicle; and the lack of sufficient fuel to power a flight RTG. The award letter also postponed launch from December 2004 to January 2006, which implied a nominal 5 year delay in arrival date from mid-2012 to mid-2017. NASA also soon insisted on New Horizons using the more expensive RTG of the two in inventory. Also complicating matters was the tragic loss of two key APL engineers responsible for REX ultrastable oscillator development in a small aircraft accident at the end of 2001.

The New Horizons team nonetheless began work in January 2002, initially focusing on the requirements development and documentation phase that would lead to a May 2002 System Requirements Review (SRR). PI Stern and the mission design team worked hard to shorten the flight time and move the arrival date earlier than 2017, ultimately achieving a mid-2015 date.

The science team, the larger planetary science community, APL, SwRI, and others worked to see funds included in the FY03 budget for mission development after Phase B. A key aspect of this battle was meeting

Dr. Weiler's challenge that NASA and the Administration would support New Horizons if the soon-to-be finalized NRC Decadal Report in Planetary Sciences (Belton et al. 2002) ranked PKB as the highest priority new start for solar system exploration. Owing to the scientific significance of the Kuiper belt exploration in general and Pluto system exploration in particular, this was accomplished in the summer of 2002, thereby largely ending funding battles over the mission (though severe cash flow difficulties persisted into FY03).

Major milestones in the development of New Horizons were:
- May 2002: Systems Requirements Review
- October 2002: Mission Preliminary Design Review
- July 2002: Selection of the Boeing STAR-48 upper stage
- March 2003: Non-Advocate Review and Authorization for phase C/D
- July 2003: Selection of the Atlas V 501 launch vehicle
- October 2003: Mission Critical Design Review
- May 2004: Spacecraft structure complete
- September 2004: First instrument payload delivery
- January 2004: Spacecraft structure complete
- March 2005: Final instrument payload delivery
- April 2005: Spacecraft integration complete
- May 2005: Beginning of spacecraft environmental testing
- September 2005: Spacecraft shipment to the launch site in Florida
- December 2005: Spacecraft mating with its launch vehicle
- January 2006: Launch

During the course of the development of New Horizons, both the spacecraft and instrument payloads evolved in many ways. The most important spacecraft changes during development included:
- RTG fuel production difficulties that resulted in a 30 watt (15%) power decrease at Pluto from design planning
- RTG mount and spacecraft balance issues that added over 50 kg in dry mass
- Downsizing of the telecom high gain antenna from 3.0 to 2.1 m to save mass
- A 25% increase in the power system capacitor bank to handle power transients up to 33 milli-Farads
- Removing corners on the triangular spacecraft structure
- Increasing the onboard solid state memory to 64 Gbit
- Substituting heavier sun trackers when advanced development units stalled in production
- Substituting traveling wave tubes for solid state power amplifiers in the telecom system
- Changes in thruster positioning to accommodate plume impingement and fuel line routing concern
- Added telecom redundancy through cross strapping of the antenna and receiver/transmitter networks.

The most important instrument payload changes during development included:
- Addition of the EPO Student Dust Counter to the payload
- Separating the PERSI instrument into distinct Ralph and Alice instruments
- Separating the PAM instrument into distinct SWAP and PEPSSI instruments
- Adding launch doors to PEPSSI, SWAP, and LORRI.

Some notable scientific developments that occurred during mission development included:
- The discovery of Kuiper belt satellites

- The discovery of factor of two increases in pressure and changes in the vertical structure of Pluto's atmosphere since 1988
- The discovery of ethane on Pluto and ammonium hydrates on Charon
- The discovery of high albedos and Pluto-like surface compositions on some KBOs
- The discovery of Pluto's satellites Nix and Hydra
- The discovery of objects as roughly large or larger than Pluto in the KB and inner Oort Cloud

During its development, over 2500 individuals worked directly on spacecraft, payload, ground system, RTG, and launch vehicle/upper stage development. Also during development, numerous personnel and programmatic changes also took place. The initial New Horizons project manager, Mr. Tom Coughlin, retired. Tom was replaced by APL's highly qualified Mr. Glen Fountain at the start of 2004. Our initial project scientist, Dr. Andy Cheng stepped down during development by the amply qualified Dr. Hal Weaver to permit Dr. Cheng to place more emphasis on his critical role as the LORRI instrument PI. It is worth noting, to APL's credit, that in both of these cases, APL followed the express choices of the PI for the replacement personnel. The NASA Marshall Space Flight Center Discovery-New Frontiers Program office came into being in late 2004, following after the dissolution of the JPL Discovery-New Frontiers Program office that operated from late 2003 to mid-2004.

These and other details of the development of New Horizons could easily fill an entire book; they even might some day. For now, the above listed summary will suffice as an introduction to the papers that follow in this volume of Space Science Reviews.

## 4. Launch and Early Flight

On schedule as directed at selection in November 2001, New Horizons took flight in January 2006. It launched on its third countdown launch attempt, at 1900 UT on 19 January 2006. This date, coincidentally, was five years to the day since the PKB AO was released, 10 years to the week since the death of Pluto's discoverer, Clyde Tombaugh, and 76 years to the week since the discovery images of Pluto were obtained. Previous launch attempts on 17 and 18 January 2006 had been foiled by a weather front that adversely affected the launch site near Cape Canaveral, Florida on 17 January and the APL mission operations site in Columbia, Maryland on 18 January. It is worth noting that 19 January was the first date on which the Atlas V actually attempted to count to zero, and when it did, it launched.

At launch, New Horizons carried all of the instruments proposed for its scientific payload in 2001 as well as the Student Dust Counter (SDC) added by PI Stern as an EPO enhancement in 2002. When launched, New Horizons carried 78 kg of fuel and pressurant; virtually all of the 80 kg load it could possibly have carried in its fuel tank. The spacecraft also carried nine mementos to its target of the ninth planet. These were: two U.S. flags, the state quarters of Maryland and Florida, a small piece of the first private manned spacecraft, SpaceShip 1, a CDROM with over 100,000 names being sent to Pluto, another CDROM with numerous pictures of the spacecraft and spacecraft-mission development teams, a 1990 US postage stamp ("Pluto: Not Yet Explored"), and a small amount of the ashes of Pluto's discoverer, Clyde Tombaugh, whose remains have become the first of a human being on their way to the stars.

The New Horizons Atlas V launch vehicle and STAR-48 upper stage both performed flawlessly, releasing the spacecraft on the proper trajectory some 50 minutes after launch. The spacecraft was contacted three minutes later and was found to be in good health. Subsequent tracking revealed the spacecraft to have received a highly accurate injection onto its Jupiter course, with velocity dispersions of order only 18 meters/second. This was far less than the ~100 meter/second fuel budget New Horizons carried for the purpose of post-launch trajectory correction. As a result, the spacecraft fuel supply available for mission science at Pluto and to explore KBOs is almost twice as large as nominal preflight predictions.

New Horizons was the fastest spacecraft ever launched. It crossed the orbit of the moon in ~9 hours and reached Jupiter in record time—just 13 months. It will reach Pluto 9.5 years after launch and will then continue across the Kuiper belt on a hyperbolic trajectory that will escape to interstellar space.

During the first 10 weeks of flight, the spacecraft was spun down to its nominal cruise 5 RPM spin rate, and a thorough series of subsystem checkouts was conducted. These subsystem tests revealed very good performance in all subsystems. No significant hardware problems were revealed on the spacecraft. Also during this period a series of three trajectory correction maneuvers were carried out to refine the course to Jupiter; 20 meters/second of fuel were expended to carry out this sequence of maneuvers.

During the period from late March to late September 2006, all seven of the payload instruments were turned on, checked out, and calibrated. All of the instruments are working well. Six of the seven instrument door deployments were completed in this phase; the last, opening the ALICE UVS solar occultation port to space was planned to take place after the spacecraft passes 4.5 AU. No instrument problems are known to exist that will compromise their scientific capabilities at the Pluto system or KBOs, though some minor annoyances have been documented in PEPSSI and RALPH.

Also during March-September 2006, several significant spacecraft software upgrades were performed. These added significant new capability to the onboard fault detection and correction ("autonomy") system, and corrected various software bugs and idiosyncrasies in both the Command and Data Handling (C&DH) and Guidance and Control (G&C) software package that had been detected after launch.

On 13 June 2006, New Horizons serendipitously flew past a small (~4 km diameter), S-type asteroid called 2002 JF56 at a fortuitously small range of just 104,000 kilometers. This "encounter," while distant from a scientifically important range, did result in some insights into this small body (Olkin et al. 2006). More importantly, however, the flyby was used to test instrument pointing and image motion compensation capabilities with a moving target. These tests were highly successful.

New Horizons encountered Jupiter on 28 February 2007. Surrounding this event, from January to June 2007, New Horizons will conduct an extensive series of Jupiter system observations. The Jupiter encounter was planned out to further calibrate instruments, test spacecraft and ground system procedures and capabilities as a risk reduction in advance of the Pluto encounter, and obtain new science to follow up on discoveries made by the most recent Jupiter missions: Galileo and Cassini. More data is expected to be generated and downlinked from Jupiter than at Pluto, thanks to the higher New Horizons telemetry rates available at ~5 AU than ~32 AU.

New Horizons will require almost 8 years of cruise to fly from the end of its Jupiter encounter to the start of its Pluto encounter. During this time the spacecraft will spend 10 months of each year in hibernation, preserving avionics lifetimes by shutting down much of its equipment and reducing costs by relaxing the need for constant monitoring and commanding. Each week, the spacecraft operators will check a beacon tone indicating general health status from the spacecraft. Once each month, engineering telemetry will be collected to assess spacecraft health and subsystem trends in more detail. For two months each year, the spacecraft will be awakened for thorough check outs, instrument calibrations, cruise science, and trajectory corrections (if necessary). In 2012 and 2014, the spacecraft and ground team will additionally conduct full-scale Pluto encounter rehearsals in flight. Plans are in place for further flight software upgrades in 2008-2009 based on lessons learned at Jupiter.

The Pluto system encounter will span ~200 days beginning in early 2015 and lasting until about 30 days after Pluto closest approach. Complete data transmission, however, will require some 4 to 9 months after the encounter owing to the 1000 bit/sec downlink rate at Pluto.

Near closest approach, RALPH will obtain maps of Pluto and Charon with km-scale resolution; at closest approach, LORRI images at scales as high as 25 m/pixel may be achieved (depending on the final flyby distance selected). In addition, the Group 1 objectives call for mapping the surface composition and distributions of major volatile species, for which RALPH will obtain: (i) four-color global (dayside) maps at 1.6 km resolution, (ii) diagnostic, hyper-spectral near-infrared maps at 7 km/pixel resolution globally (dayside), and at 0.6 km/pixel for selected areas. Maps of surface-lying $CH_4$, $N_2$, CO, $CO_2$, and $H_2O$ abundances will be obtained. Surface temperatures will be mapped by RALPH using temperature-sensitive IR spectral features; these maps are expected to have resolutions as good as 2 deg K and 10 km; hemispheric-averaged surface brightness temperature will also be measured by the REX radiometer mode.

Characterization of Pluto's neutral atmosphere and its escape rate will be accomplished by: (i) ALICE ultraviolet airglow and solar occultation spectra to determine the mole fractions of $N_2$, $CH_4$, CO and Ar to 1% in total mixing ratio and to determine the temperature structure in the upper atmosphere, (ii) REX radio occultations at both Pluto and Charon, measuring the density/temperature structure of Pluto's neutral atmosphere to the surface, (iii) SWAP and PEPSSI in situ determination of the atmospheric escape rate by measuring Pluto pickup ions, and (iv) ALICE H Lyα mapping of Pluto and Charon in order to determine the rate of Roche-lobe flow of atmosphere from Pluto to Charon. Searches for atmospheres around Charon and KBOs will be made using ALICE with both airglow and solar occultation techniques.

REX-derived Doppler tracking will also be used to measure the masses of Pluto and Charon, and to attempt J2 determinations; together with imagery-derived 3-D volumes, these data will be used to obtain improved densities. SDC will measure the density and masses of dust particles in the solar system from 1 AU to at least 40 AU, far surpassing the 18 AU boundary beyond which any dust detector has as yet penetrated.

Approximately three weeks after the Pluto closest approach, New Horizons will perform a trajectory correction maneuver to target its first KBO encounter. The target KBO will be selected after a deep search for candidates to be made in 2011-2013 after Pluto exits the dense background star fields of Sagittarius. Monte Carlo simulations indicate that 4 to perhaps 12 candidate KBOs may be detected within reach of the spacecraft's available fuel supply. These same simulations indicate a typical 2 to 3 year cruise time to reach the first KBO flyby. A second KBO flyby may also be possible given the available project fuel. It is important to note, however, that any KBO encounter will require the approval of an extended mission budget from NASA.

## 5. Conclusion

New Horizons is now safely in flight to make the first reconnaissance of the fascinating Pluto-system: the farthest of the classical planets, the most accessible example of ice dwarf planets, and the most well known KBO in 2015.

If all goes well, New Horizons will then go on to make a flyby reconnaissance of a 30-50 kilometer diameter KBO some two to four years later. A second KBO encounter two to four years later still may even be possible, depending on spacecraft health, fuel status, our ability to detect sufficient KBOs within reach of the vehicle before it passes out of the classical KB at about 50 AU, and NASA funding.

Beyond completing the initial reconnaissance of the nine classical planets, New Horizons has already demonstrated the ability of PI-led missions to be mounted to the outer solar system, opening up a wide range of future possibilities. As well, the mission has proven the ability of non-traditional entities, like APL and SwRI to explore the outer solar system, giving NASA new programmatic flexibility and enhancing the competitive options available to NASA when selecting outer planet missions.

If New Horizons is successful, it will represent a watershed development in the scientific exploration of dwarf planets—an entirely new class of bodies in the solar system, of worlds with exotic volatiles on their surfaces, of rapidly (possibly hydrodynamically) escaping atmospheres, and of giant impact derived satellite systems. It will also provide the first dust density measurements beyond 18 AU, cratering records that shed light on both the ancient and present-day KB impactor population down to 10s of meters, and a key comparator to the puzzlingly active, former dwarf planet (now satellite of Neptune) called Triton, which is as large as Eris and Pluto.

Following this article are a series of other articles describing in significantly more detail the mission trajectory, the mission science, the spacecraft, the science payload and its individual instruments, and the mission EPO program.

## Acknowledgements

The author thanks the entire New Horizons team, including the Headquarters and NASA Marshall Space Flight Center program offices, all of the mission subcontractors, Lockheed-Martin and Boeing, the New Horizons science team, the Department of Energy, JPL, and NASA Kennedy Space Flight Center for their dedication, commitment, and drive throughout the development of this incredible mission of discovery. He also thanks New Horizons Program Scientist D. Bogan and New Horizons Project Scientist H. Weaver for helpful comments on this manuscript.

## References

Belton, M.J., et al., 2002. New Frontiers in the Solar System. An Integrated Exploration Strategy. National Research Council, 145 pp.

Farquhar, R. and Stern, A., Pushing Back the Frontier: A Mission to the Pluto-Charon System, The Planetary Report, **10**, No. 4, July-August 1990, pp. 18-23.

Lunine, J.I., et al. Report of the Pluto-Kuiper Express Science Definition Team, NASA, 1996.

NASA, 2001. Pluto Kuiper belt Mission Announcement of Opportunity. AO 01-OSS-01.

Olkin, C.O., D. Reuter, A. Lunsford, R.P. Binzel, and S.A. Stern. The New Horizons Distant Flyby of Asteroid 2002 JF56. American Astronomical Society, Division of Planetary Science meeting **38**, abstract 9.22, 2006.

Stern, S.A., 2002. Scientific American, Journey to the Farthest Planet. 286, 56-59.

Stern, S.A., and A. Cheng, 2002. NASA Plans Pluto-Kuiper Belt Mission. EOS, **83**, 101-106.

Stern, S.A., and J. Mitton, Pluto and Charon: Ice Worlds on the Ragged Edge of the Solar System, Wiley-VCH, 244 pp, 2005.

Stern, S.A., D.C. Slater, W. Gibson, H.J. Reitsema, A. Delamere, D.E. Jennings, D.C. Reuter, J.T. Clarke, C.C. Porco, E.M. Shoemaker, and J.R. Spencer, "The Highly Integrated Pluto Payload System (HIPPS): A Sciencecraft Instrument for the Pluto Mission," in *EUV, X-Ray, and Gamma-Ray Instrumentation for Astronomy VI*, Proceedings of SPIE Vol. **2518**, O.H.W. Siegmund and John Vallerga, Editors, 39-58, 1995.

Terrile, R.J., S.A. Stern, R.L. Staehle, S.C. Brewster, J.B. Carraway, P.K. Henry, H. Price, and S.S. Weinstein. "Spacecraft Missions to the Pluto and Charon System," in *Pluto and Charon*, (S.A. Stern & D.J. Tholen, eds.). University of Arizona Press, Tucson, TBD-TBD, 1997.

# Figures and Captions

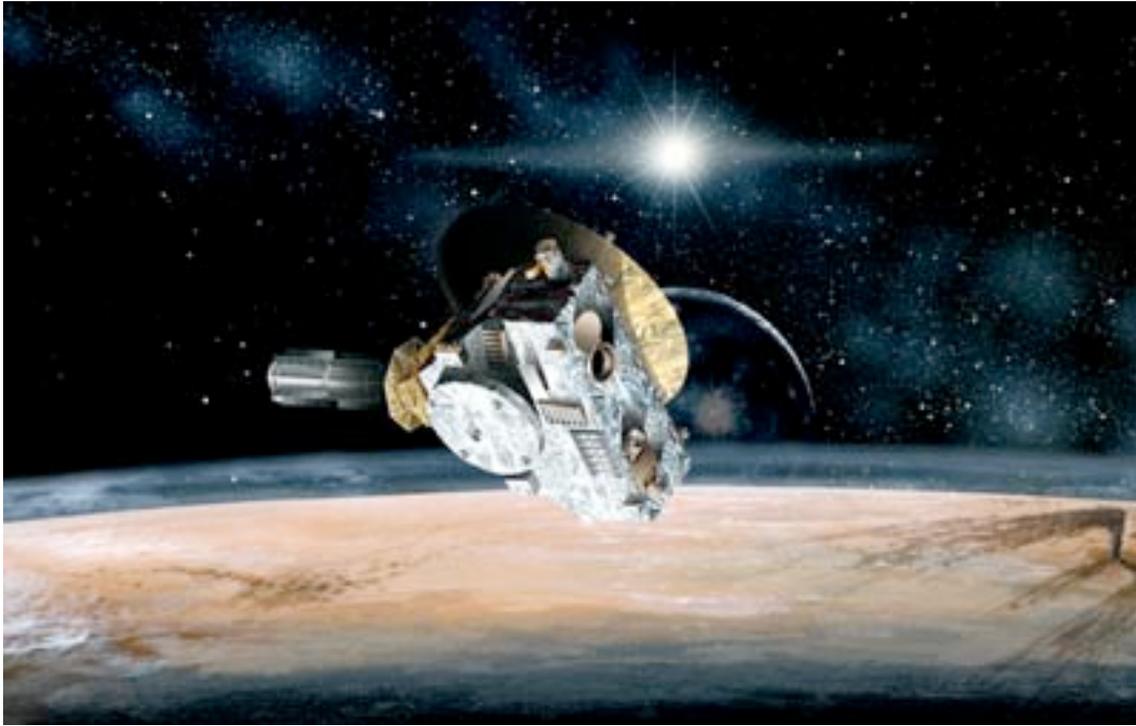

Figure 1. New Horizons at Pluto as depicted over Pluto by planetary scientist and space artist Dan Durda.

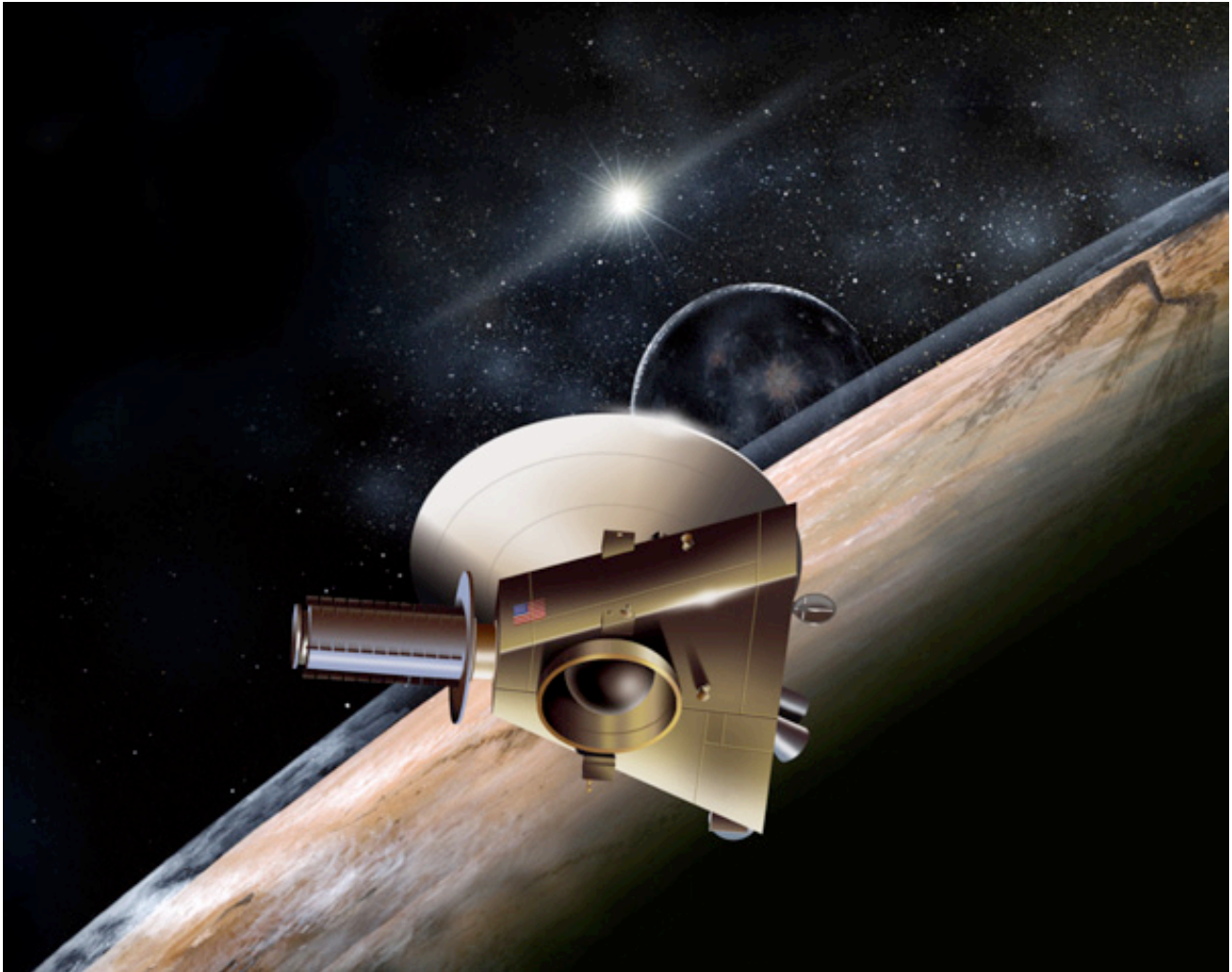

Figure 2: New Horizons spacecraft concept as originally proposed, with artwork by Dan Durda.

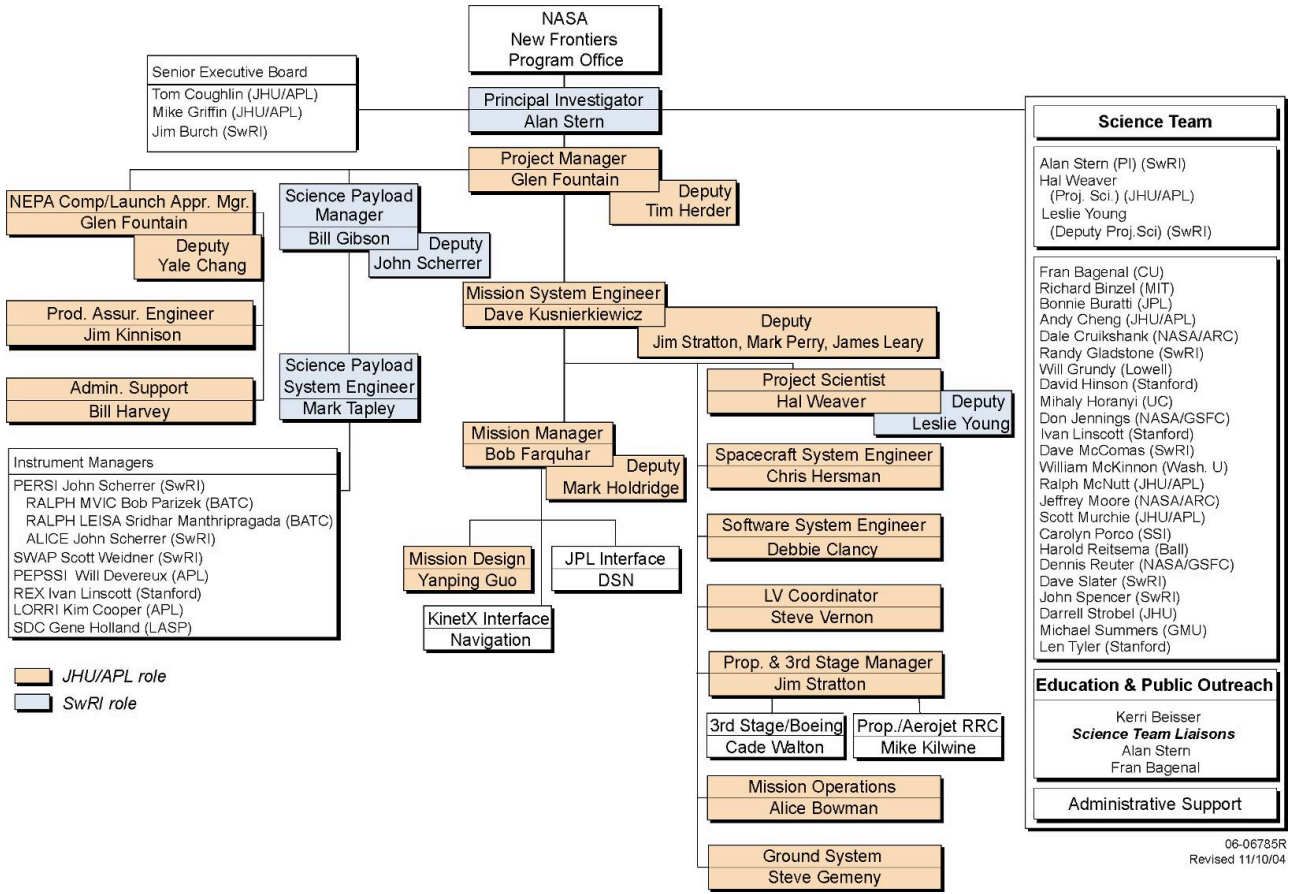

Figure 3: New Horizons spacecraft-payload team project organization chart during development.

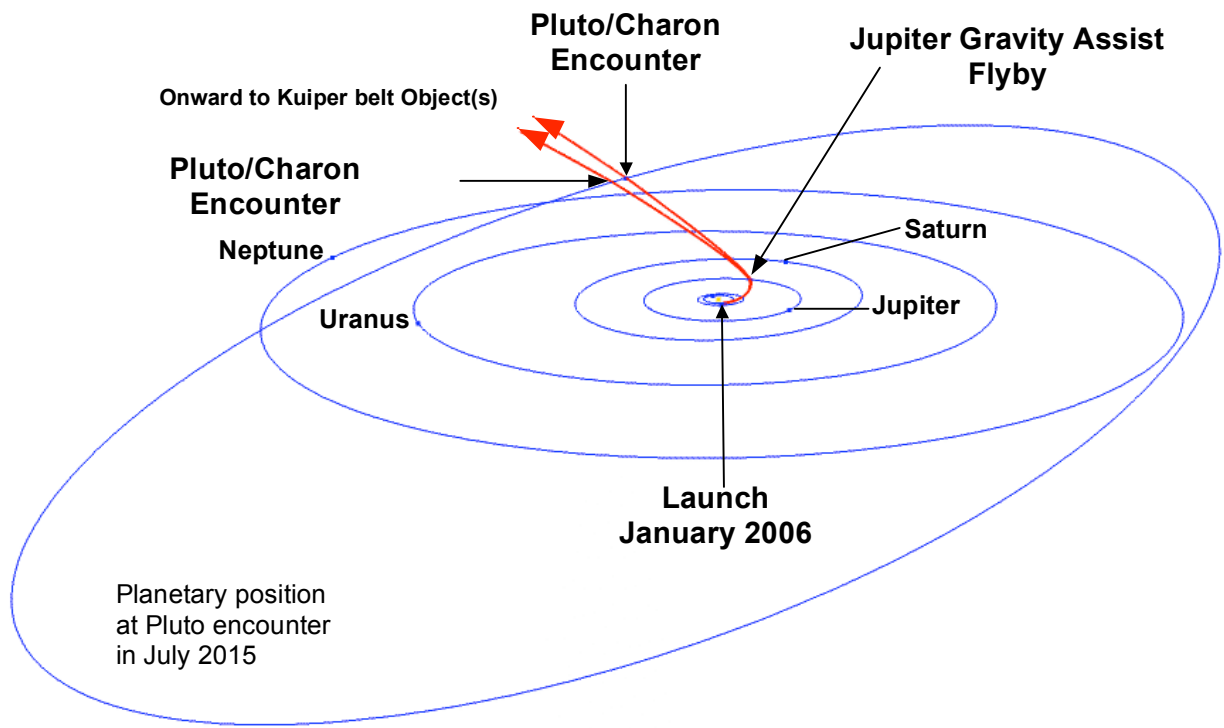

Figure 4. New Horizons trajectory depiction; the two red trajectory lines show the range of possible encounter dates (2015-2020) that applied for all possible launch dates in the 35-day long 2006 launch window. Planetary positions are shown at the time of launch.